\documentclass[aps,prb,twocolumn,superscriptaddress,longbibliography]{revtex4-2}
 
\usepackage{amsmath,amssymb}
\usepackage{bm}
\usepackage{graphicx}
\usepackage{subfigure}
\usepackage{color}
\usepackage[version=4]{mhchem}
\usepackage{wasysym} 

\begin{document}
\title{Clusterization transition between cluster Mott insulators  \\  
on a breathing Kagom\'{e} lattice}     
\author{Xu-Ping Yao}
\thanks{These authors contributed equally.}
\affiliation{Department of Physics and HKU-UCAS Joint Institute for Theoretical and Computational Physics at Hong Kong, The University of Hong Kong, Hong Kong, China}
\author{Xiao-Tian Zhang}
\thanks{These authors contributed equally.}
\affiliation{Department of Physics and HKU-UCAS Joint Institute for Theoretical and Computational Physics at Hong Kong, The University of Hong Kong, Hong Kong, China}
\author{Yong Baek Kim}
\affiliation{Department of Physics, University of Toronto, Ontario M5S 1A7, Canada}
\author{Xiaoqun Wang}
\affiliation{School of Physics and Astronomy, Tsung-Dao Lee Institute, 
Shanghai Jiao Tong University, Shanghai 200240, China}
\affiliation{Key Laboratory of Artificial Structures and Quantum Control of MOE,
Shenyang National Laboratory for Materials Science, Shenyang 110016, China}  
\author{Gang Chen}
\affiliation{Department of Physics and HKU-UCAS Joint Institute for Theoretical and 
Computational Physics at Hong Kong, The University of Hong Kong, Hong Kong, China}
\affiliation{State Key Laboratory of Surface Physics and Department of Physics, 
Institute of Nanoelectronics and Quantum Computing, 
Fudan University, Shanghai, 200433, China}
\affiliation{Collaborative Innovation Center of Advanced Microstructures, 
Nanjing University, Nanjing 210093, China}
\affiliation{Institute of Nanoelectronics and Quantum Computing, 
Fudan University, Shanghai 200433, China}

\date{\today}

\begin{abstract}
Motivated by recent experimental progress on various cluster Mott insulators, 
we study an extended Hubbard model on a breathing Kagom\'{e} lattice 
with a single electron orbital and $1/6$ electron filling. 
Two distinct types of cluster localization are found in the cluster Mott regime 
due to the presence of the electron repulsion between neighboring sites, 
rather than from the on-site Hubbard interaction in the conventional Mott insulators. 
We introduce a unified parton construction framework to accommodate both type of cluster Mott insulating phase
as well as a trivial Ferm liquid metal and discuss the phase transitions in the phase diagram.
It is shown that, in one of the cluster localization phases,
the strong inter-site repulsion results into locally metallic behavior 
within one of two triangular clusters on the breathing Kagom\'{e} lattice. 
We further comment on experimental relevance to existing Mo-based cluster magnets.
\end{abstract}

\date{\today}

\maketitle
 
\section{Introduction}

Cluster Mott insulators (CMIs) seem to become a new frontier 
for exploring the emergent correlated physics~\cite{Gang2014,Gang2018,Gang2017}. 
The cluster magnet 1T-\ce{TaS2} develops a commensurate 
charge density wave order at about 120K, and the enlarged unit cell due to the  
charge order then has a David-star shape with 13 lattice sites~\cite{Law2017,Gang2018B}. 
The enlarged unit cell traps one unpaired electron that is Mott localized on the cluster unit 
of the David star, and the system in the commensurate charge density wave  
state forms a CMI in two spatial dimensions. The surging field of twistronics,
that was initiated from the twisted bilayer graphenes~\cite{Cao2018,Cao2018B,Bistritzer2011,Xu2018,PhysRevLett.122.086402,Tang:2020bb}, 
potentially can be another example of realizing CMIs where the electrons are localized on 
the large moir\'{e} unit cell.    
These moir\'{e} unit cells are often one to two orders of magnitude larger than 
the lattice constant of the original untwisted crystals. The twisting procedure 
provides a new knob to tune the physical properties of the underlying systems. 
The common ingredient shared by these systems is the large cluster unit for the 
electronic degrees of freedom, and the longer range interactions ought to be 
considered~\cite{Gang2014,Gang2016,Gang2018,Gang2015,Gang2017,Khomskii2004,khomskii2020orbital,komleva2020threesite,
Cava2013,Streltsov_2018,Kim2014,Streltsov_2017,Radaelli2002}. This ingredient leads to distinct and interesting features and experimental consequences in different realizations of cluster localization.

%-----------------------------
\begin{figure}[b]
{\includegraphics[width=8cm]{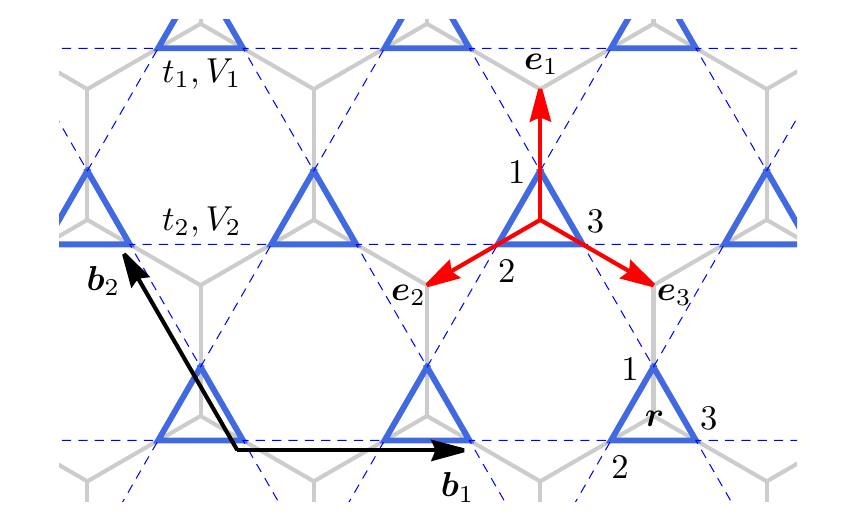}}
\caption{ Breathing Kagom\'{e} lattice.}
\label{fig1}
\end{figure}
%-----------------------------

In expectation of the large internal electronic degrees of freedom inside each cluster,
we explore the rich phase diagram of the CMIs.
We present the observation by showing the existence of distinct cluster  
localizations and study the phase transition between the CMIs on the breathing Kagom\'{e} lattice.
This is partly motivated by the experiments on various Mo-based two-dimensional 
cluster magnets~\cite{Sheckelton2012,Broholm2014,Quilliam2018,Sheckelton2015,McQueen2014}.    
We study a $1/6$-filled extended Hubbard model with the nearest-neighbor 
repulsions on a breathing Kagom\'{e} lattice. 
The Mott insulating physics in this partially filled system arises
 from the large nearest-neighbor 
repulsions~\cite{Gang2014,Gang2016,Wang2009} 
and localization of the electrons in the triangular cluster units. 
Due to the asymmetry between the up- and down-triangles  
and the resulting difference in the interactions and hoppings,   
two different cluster localizations are expected.                     
We first show that, for the case of cluster localization on 
only one type of triangular units (e.g. the up ones) in the strong breathing limit, 
the ground state is smoothly connected to the one for the triangular lattice  
Hubbard model at half filling. We then explore the phase transition between 
two distinct cluster localizations and further address the consequences 
on the spin physics. In terms of the lattice gauge theory formulation, 
this transition is identified as a Higgs transition. The correspondence between 
the lattice gauge theory formulation and the physical variables are clarified. 

The remainder of the paper is organized as follows. We begin in Sec.~\ref{secII} by introducing the extended Hubbard model on the breathing Kagom\'{e} lattice with a single electron orbital and $1/6$ electron filling. In Sec~\ref{secIII} we explore the strong breathing limit and discuss the type-I CMI and the Mott transition. We go beyond the strong breathing limit in Sec.~\ref{secIV} and study the type-II CMI as well as its emergent U(1)$_{\text{c}}$ gauge structure. We further introduce a unified parton construction in Sec.~\ref{secV} to reveal the rich physics of both type-I and type-II CMIs at the mean-field level. We establish the generic phase diagram and discuss the phase transition between two distinct CMIs. We conclude by discussing the experimental relevance and consequence about Mo-based cluster magnets in Sec.~\ref{secVI}.

\section{Extended Hubbard Model}
\label{secII}

We start from the extended Hubbard model on the breathing Kagom\'{e} lattice,
\begin{eqnarray}
H &=&  \sum_{ \langle ij \rangle  } 
\big[ {-t_{ij}^{}}  c^{\dagger}_{i\sigma} c^{\phantom\dagger}_{j\sigma}  + V_{ij} n_i n_j 
\big]
+   \sum_{i} U n_{i\uparrow} n_{i\downarrow},
\label{eq1}
\end{eqnarray}
where
$c^{\dagger}_{i\sigma}$ ($c^{\phantom\dagger}_{i\sigma}$) creates  
(annihilates) an electron with spin $\sigma$ at the lattice site $i$, and 
${t_{ij} = t_1 (t_2)}, {V_{ij} = V_1 (V_2)}$ for $ij$ on the up- (down-) triangles. 
Here ${n_i \equiv \sum_{\sigma} c^\dagger_{i\sigma}c^{\phantom\dagger}_{i\sigma}}$ 
defines the electron occupation number at the lattice site $i$. We 
are interested in the regime with one electron in each Kagom\'{e}  
lattice unit cell, and thus the electron filling for this Hubbard model 
is $1/6$~\cite{Sheckelton2012,Quilliam2018}. 
This model is expected to capture the essential physics of 
the Mo-based cluster magnets~\cite{Gang2016,Gang2018}. 
For the fractional $1/6$ filling here, 
the Mott localization is driven by the inter-site repulsions ($V_1, V_2$) 
rather than the on-site Hubbard-$U$ interaction and the electrons 
are localized in the (elementary) triangles of the Kagom\'{e} 
lattice instead of the lattice sites. Due to the asymmetry between 
the up- and down-triangles, the Mott localization in the up-triangles
and down-triangles does not occur simultaneously. 
Setting the Hubbard-$U$ as the largest energy scale, we study 
the properties of the model and explore its phase diagram.

The kinetic part of the model can be readily diagonalized 
and the electrons form the following three bands, 
\begin{eqnarray}
E_{1,2} ( {\bf k}) & =& 
- \frac{1}{2} \big[ t_1 +t_2  \pm [9t_1^2 -6t_1 t_2  + 9t_2^2 \nonumber \\
&+& 8t_1 t_2 \big( \cos k_1 + \cos k_2 + \cos (k_1+k_2) \big) ]^{ \frac{1}{2}}  \big], 
\\
E_3({\bf k})  &=& t_1 + t_2, 
\end{eqnarray}
where ${k_1 \equiv {\bf k}\cdot {\bf b}_1},{ k_2 \equiv {\bf k}\cdot {\bf b}_2 }$,
and ${\bf b}_1$, ${\bf b}_2$ are two elementary lattice vectors
of the underlying Bravais lattice (see Fig.~\ref{fig1}). These 
three electron bands are well-separated from each other 
and only touch at certain discrete momentum points.   
In particular, ${E_{1} ( {\bf k})}$ and ${E_{2} ( {\bf k})}$ 
have Dirac-point band touchings at the Brillouin zone corners 
when ${t_1 = t_2}$. With the $1/6$ electron filling, the electrons  
fill half the lowest band ${E_1 ({\bf k})}$ and the   
ground state of the kinetic part is a Fermi liquid (FL) metal. 

\section{Strong Breathing Limit and Type-I CMI}
\label{secIII}

Electron correlations are considered on top of the 
kinetic part. A strong Hubbard-$U$ merely suppresses 
double occupation on a single lattice site and cannot 
cause localization due to the fractional filling here. What replaces 
 is the cluster localization from the inter-site interactions.
We first explore the strong breathing limit with ${V_2=0}$ 
and study the cluster localization driven by the remaining 
interaction $V_1$ in the framework of the slave-rotor 
construction~\cite{Georges2004,Lee_2005,Lee_2005b,Lee2012}. 
A strong repulsion $V_1$ penalizes the double occupancy 
on the up-triangles and would drive a Mott transition from 
FL metal to a CMI. Because the number of 
the up-triangles is equal to the total electron number, 
there is exactly one electron in each up-triangle in this cluster 
Mott regime. To describe different phases and study the Mott transitions, 
we first employ the standard slave-rotor representation for 
the electron operator
${c^\dagger_{i\sigma} = f^\dagger_{i\sigma} e^{i \theta_i}}$,
where the bosonic rotor ($e^{i\theta_i}$) carries the electron charge and
the fermionic spinon ($f^\dagger_{i\sigma}$) carries the spin quantum number.  
To constrain the enlarged Hilbert space, we introduce an angular momentum 
variable $L_i^z$, 
\begin{eqnarray}
{L_i^z = \sum_{\sigma} f^\dagger_{i\sigma} 
f^{\phantom\dagger}_{i\sigma}  - {1}/{2}}, 
\end{eqnarray}
where $L_i^z$ is conjugate to the rotor variable with 
\begin{eqnarray}
{[\theta_i,L_j^z] = i \delta_{ij}}. 
\end{eqnarray}
Since the interaction $U$ is assumed to be the largest,
in the large $U$ limit the double electron occupation is always suppressed. 
Hence, the angular variable $L_i^z$ primarily takes ${L_i^z = {1}/{2}}$ 
($-{1}/{2}$) for a singly-occupied (empty) site. Decoupling
the electron hopping into the spinon and the rotor sectors, 
we obtain the Hamiltonians for the spin and charge sectors
\begin{eqnarray}
H_{\text{s}} &=& -\sum_{\langle ij \rangle} \tilde{t}_{ij} 
f_{i\sigma}^{\dagger} f_{j\sigma}^{} -h \sum_i 
f^{\dagger}_{i\sigma} f^{}_{i\sigma},
\\
H_{\text{c}} &=& -  \sum_{\langle ij \rangle   } 
          2 J_{ij} \cos( \theta_i - \theta_j)  
+  \frac{V_1}{2} 
\sum_{{\bf r} }\mathbb{L}_{\bf r}^2  
 \nonumber  \\
&&  
+ (h + \frac{5V_1}{2}) \sum_{ {\bf r}   }  \mathbb{L}_{\bf r}
+ \frac{U-V_1}{2} \sum_i  L_i^2 ,
\label{eq3}
\end{eqnarray}
where ${\bf r}$ labels the center of up-triangle or equivalently
the unit cell of the lattice, 
\begin{eqnarray}
\tilde{t}_{ij} &=& t_{ij} \langle e^{i \theta_i - i\theta_j} \rangle \equiv |t_{ij}|e^{ia_{ij}} ,\\
J_{ij} &=& t_{ij} \sum_{\sigma} \langle f^{\dagger}_{i \sigma} f^{}_{j\sigma}  \rangle 
\equiv |J_{ij}| e^{-ia_{ij}}, 
\end{eqnarray}
and $h$ is a Lagrangian multiplier to enforce the Hilbert space constraint. 
The Hamiltonian is invariant under an internal U(1) gauge transformation, 
${f^{\dagger}_{i\sigma} \rightarrow f^{\dagger}_{i\sigma}e^{-i\chi_i} }$,
${\theta_i \rightarrow \theta_i +\chi_i}$, and ${a_{ij} \rightarrow a_{ij} +\chi_i-\chi_j}$.
Here we have introduced an angular momentum 
operator $\mathbb{L}_{\bf r} $ as 
\begin{eqnarray}
  \mathbb{L}_{\bf r} &  = & 
  [\sum_{\mu} L_{{\bf r} \mu}]   + {1}/{2}  \equiv   \sum_{\mu,\sigma}  
                     f^{\dagger}_{{\bf r}\mu\sigma}f^{\phantom\dagger}_{{\bf r}\mu\sigma} - 1 ,
\end{eqnarray}
 where the lattice site is labeled by the combination of the 
 unit cell ${\bf r}$ and the sublattice index $\mu$.  $\mathbb{L}_{\bf r}$ 
 measures the total electron occupation on the up-triangle at ${\bf r}$. 
 Moreover, from $L_{{\bf r}\mu}$, we find that $\mathbb{L}_{\bf r}$ can take $-1,0,1,2$. 
 Finally, because ${L_i^z = \pm 1/2}$, the last term in the second line of Eq.~\eqref{eq3} 
 reduces to a constant and can be dropped. It is convenient to define the conjugate 
 variable for $\mathbb{L}_{\bf r}$. We introduce a super-rotor operator 
 $e^{\pm i \Theta_{\bf r}}$ whose physical meaning is to create and annihilate 
 an electron charge in the up-triangle at ${\bf r}$.  Clearly, we have 
\begin{equation}
\Theta_{\bf r}  \equiv  \frac{1}{3} \sum_{\mu} \theta_{{\bf r} \mu}, \quad\quad
[\Theta_{\bf r}, \mathbb{L}_{{\bf r}' }   ] = i \delta_{{\bf r}{\bf r}'}. 
\end{equation}
As ${[\theta_{{\bf r}\mu} - \theta_{ {\bf r}\nu},  \mathbb{L}_{\bf r'}] = 0} $, 
the hopping terms \emph{inside} the up-triangles 
commute with the $V_1$ interaction and we can set 
$ {{\theta_{{\bf r}\mu}  \equiv \Theta_{\bf r}}}$. 
The hopping terms inside the down-triangles,
that describe the electron tunneling from the neighboring up-triangles,
do not commute with the $V_1$ interaction. 
Increasing $V_1$ penalizes the kinetic 
energy gain through hoppings on the down-triangle bonds 
and causes the electron cluster localization in the up-triangles.
Nevertheless, the electrons remain mobile inside each up-triangle and 
can gain kinetic energy through hoppings within the up-triangle. 
Thus, the system is locally ``metallic'' within each up-triangle and 
remains so even when the interaction $V_1$ becomes dominant.

%-----------------------------
\begin{figure*}[t]
\subfigure[\, $V_2 = 0$]{\includegraphics[width=7.6cm]{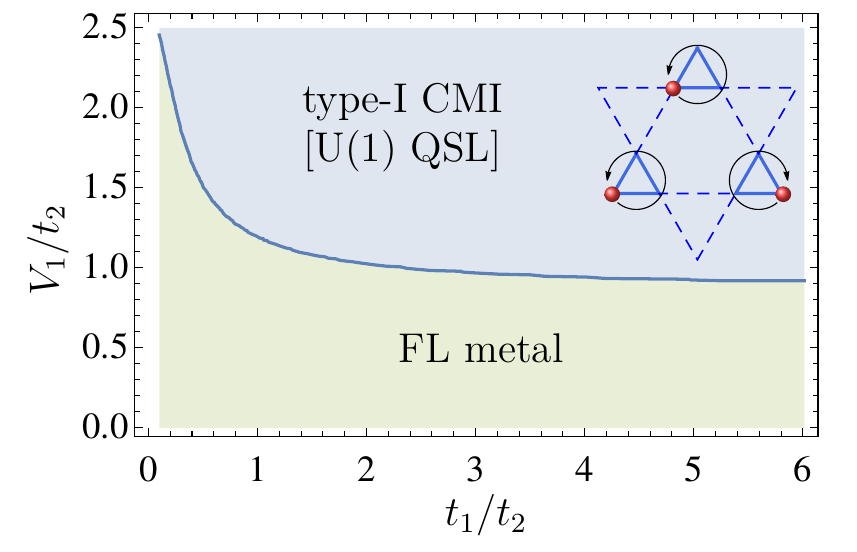}}
\subfigure[\, $t_1 = t_2$]{\includegraphics[width=5.05cm]{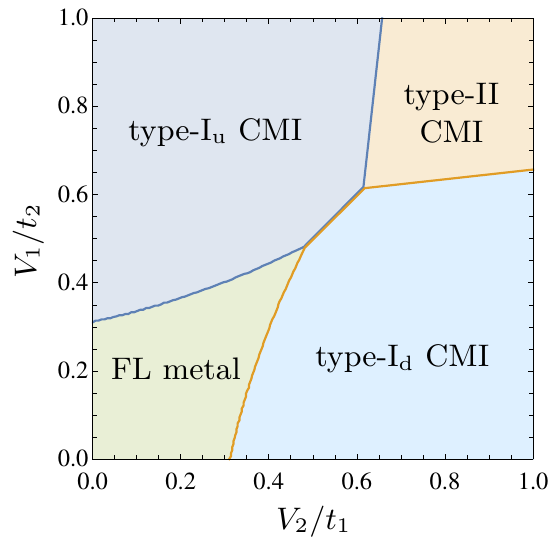}}
\subfigure[\, $t_1 =2 t_2$ ]{\includegraphics[width=5.05cm]{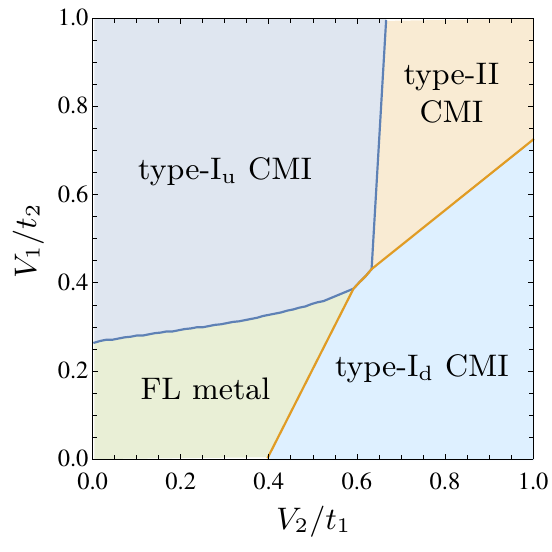}}
\caption{ (a)
The slave-rotor mean-field phase diagram at ${V_2=0}$.
We exclude the 120-degree state in the strong coupling limit (${V_1 \gg t_2}$). 
Inset describes the free and uncorrelated motion of the 
electrons inside the up-triangles, and the direction is arbitrarily chosen. 
(b,c) The phase diagram of the extended Hubbard model for different parameters. }
\label{fig2}
\end{figure*}
%-----------------------------

Using the local metallic condition ($ {{\theta_{{\bf r}\mu}  \equiv \Theta_{\bf r}}}$) 
to optimize the intra-up-triangle hopping, we obtain a reduced rotor 
Hamiltonian that is defined on the triangular lattice formed by the 
centers of the up-triangles, 
\begin{eqnarray}
\tilde{H}_{\text{c}} &=& 
- 2 {J}_2 \sum_{ \langle {\bf r} {\bf r}\rq{} \rangle } 
\cos (\Theta_{\bf r} - \Theta_{{\bf r}\rq{}} ) \nonumber \\
&& + \sum_{{\bf r}}\big[ \frac{V_1}{2} \mathbb{L}_{\bf r}^2 +\tilde{h}  \mathbb{L}_{\bf r} \big],
\end{eqnarray} 
where ${\langle {\bf r} {\bf r}' \rangle}$ labels two neighboring up-triangles 
and ${\tilde{h} = h + 5V_1/2}$, $J_2$ is defined on down-triangles. 

The relevant degrees of freedom for the Mott transition is 
the super-rotor mode $e^{i\Theta_{\bf r}}$. When it is condensed 
and ${\langle e^{i\Theta_{\bf r}} \rangle \neq 0}$, 
we obtain a FL metal. When it is gapped with 
${\langle e^{i\Theta_{\bf r}} \rangle = 0}$, a CMI 
with electrons localized on all up-triangles is obtained,
and we refer this CMI as type-I CMI in Fig.~\ref{fig2}(a).   
In this type-I CMI, there exists charge coherence within the up-triangle as it is 
``locally metallic". The gauge field fluctuations within the up-triangles become
massive from the Higgs mechanism. The gauge fluctuations on the links between
two neighboring up-triangles remain gapless and we represent it 
by $a_{{\bf r}{\bf r}\rq{}  }$ for two up-triangles at ${\bf r}$ and ${\bf r}\rq{}$.
The reduced rotor Hamiltonian $\tilde{H}_{\text{c}}$ and the spinon 
Hamiltonian $H_{\text{s}}$ are invariant under the U(1) gauge transformation
${f^\dagger_{{\bf r}\mu\sigma} \rightarrow f^\dagger_{ {\bf r}\mu\sigma} e^{- i\chi_{\bf r}}},
{\Theta_{\bf r} \rightarrow \Theta_{\bf r} + \chi_{\bf r}},
{a_{{\bf r}{\bf r}\rq{}} \rightarrow a_{{\bf r}{\bf r}\rq{}} + 
\chi_{\bf r} - \chi_{{\bf r}\rq{}}}$. 

In the type-I CMI, the spinon mean-field Hamiltonian $H_{\text{s}}$ 
describes the spinon hopping at the mean-field level.
The spinon bands are identical to the electronic ones,
${E_{\mu} ({\bf k})}$, except for the modified hopping. 
Thus, the spinons fill a half of the lowest spinon band,     
leading to a spinon Fermi surface. The resulting spin sector 
is a U(1) quantum spin liquid (QSL) with a spinon Fermi surface. 
It is generally believed that, the U(1) QSL is in the deconfined 
phase due to the spinon Fermi surface that suppresses the instanton
events. When the super-rotor mode is condensed, the U(1) gauge 
field picks up a mass via the Higgs' mechanism, and the charge rotor 
and fermionic spinons are then combined back to the original electron.
Here we solve the charge sector Hamiltonian $\tilde{H}_{\text{c}}$ and the spinon Hamiltonian $H_{\text{s}}$ self-consistently for 
the phase diagram and Mott transition. Following the standard procedure, we implement the coherent state path integral for the super-rotor variables ${\phi^{\dagger}_{\bf r} \equiv e^{i\Theta_{\bf r}}}$ 
and ${\phi_{\bf r} \equiv e^{-i\Theta_{\bf r}}}$. By integrating out the field $\mathbb{L}_{\bf r}$, we obtain the partition function, 
\begin{equation}
\mathbb{Z} = \int \mathcal{D} \phi^\dagger \mathcal{D} \phi \mathcal{D} \lambda 
e^{- \mathcal{S} -  \sum_{{\bf r} \in \text{u} } \int d\tau \lambda_{\bf r} ( | \phi_{\bf r} |^2 -1 )  },
\label{eq01}
\end{equation}
with the effective action
\begin{equation}
\label{eq02}
\mathcal{S} = \int d\tau  \sum_{{\bf r} } \frac{1}{2 V_1} |\partial_{\tau} 
\phi_{\bf r}  |^2  - 
J_2  \sum_{\langle {\bf r} {\bf r}\rq{}  \rangle } 
( \phi^\dagger_{\bf r} \phi^{\phantom\dagger}_{{\bf r}\rq{} }  + h.c. ).
\end{equation}
We have dropped the term with parameter $\tilde{h}$ that is required to vanish   
since ${\sum_{{\bf r}  } \langle \mathbb{L}_{\bf r} \rangle = 0}$. The Lagrange multiplier 
$\lambda_{\bf r}$ is also introduced in the partition function to 
enforce the unimodular constraint ${|\phi_{\bf r} | =1}$ for each up-triangle. 
We take a uniform saddle point approximation by setting ${\lambda_{\bf r} = \lambda}$ 
and further integrate out the $\phi$ fields. Finally we get the following saddle point 
equation in the Mott insulating phase, 
\begin{equation}
\frac{1}{S_{\text{BZ}}}  \int d^2 {\bf k} \frac{V_1}{ \omega_{\bf k}} =1,
\label{saddleEq}
\end{equation}
where $S_{\text{BZ}}$ is the area of the first Brillouin zone of the triangular
lattice and $\omega_{\bf k}$ is the dispersion of the super-rotor mode with  
\begin{equation}
\omega_{\bf k} = \Big[
2V_1 
\big(
\lambda - 2J_2 ( \cos k_1
+ \cos k_2 + \cos (k_1+k_2)
\big) 
\Big]^{\frac{1}{2}}.
\end{equation}
When ${\lambda = 6 J_2}$, the dispersion 
$\omega_{\bf k}$ becomes gapless. That means the super-rotor mode is condensed.
Combining this condensation condition with the super-rotor saddle point equation 
Eq.~\eqref{saddleEq} and the spinon-sector mean-field theory, we construct the phase diagram in the strong breathing limit as shown in Fig.~\ref{fig2}(a).

Here we do not consider the possibility of magnetic ordering in the 
strong Mott regime. For a small (large) $V_1/t_2$, we obtain a Fermi 
liquid metal (or a U(1) QSL with a spinon Fermi surface). The 
Mott transition is continuous and of the quantum XY type in 
the mean-field theory, and is expected to be so even after 
including the U(1) gauge fluctuations~\cite{Senthil2008A}.
The phase boundary of the Mott transition is understood as follows. 
For smaller (larger) $t_1/t_2$, the electrons gain more (less)
kinetic energy from the $t_2$ hopping or the inter-up-triangle hopping, 
and thus, a larger (smaller) critical $V_1/t_2$ is needed to localize
the electrons in the up-triangles. In particular, in the limit of 
${t_1/t_2 \rightarrow \infty}$, our model with ${V_2 = 0}$ 
and $1/6$ electron filling is equivalent to a triangular lattice Hubbard model 
at half-filling where the triangular lattice is formed by the up-triangles. 
Therefore, the U(1) QSL with a Fermi surface 
in the type-I CMI is smoothly connected to the one
proposed for the triangular lattice Hubbard model 
at half-filling~\cite{Lee_2005,Lee_2005b}. 

%%%%%%%%%%%%

\section{Emergent U(1)$_\text{c}$ gauge structure in type-II CMI}
\label{secIV}

As $V_2$ gradually increases from zero, the free motion of electrons inside 
the up-triangles becomes less favorable energetically because this motion 
creates double occupancy configurations on the down-triangles. Thus, 
at a critical $V_2$, the electron number on each down-triangle 
is also fixed to be one, and we experience the cluster localization on both 
types of triangles. This new cluster Mott state is referred as 
type-II CMI.
The slave-rotor representation in this phase 
fails to capture the proper physics and we should introduce a new parton construction. 
Before that, we need to first understand the low-energy physics of the charge sector, 
especially in the type-II CMI. We will show the charge localization pattern in the type-II CMI 
leads to an emergent compact U(1) lattice gauge theory description
 for the charge-sector quantum fluctuations. 
In the slave-rotor formalism, the charge-sector Hamiltonian with $V_2$ 
interaction is given by
\begin{eqnarray}
H_{\text{c}} & =&  
\sum_{\langle ij\rangle } 
{- 2J_{ij}}\cos (\theta_i -\theta_j) 
+ V_{ ij}( L_i^z + \frac{1}{2})  (L_j^z + \frac{1}{2}) 
\nonumber \\
&&+ \sum_i h (L_i^z + \frac{1}{2}),
\end{eqnarray}
where we have dropped the $U$ interaction term because 
${L_i = \pm 1/2}$ only gives a constant for the $L_i^2$ term in the large $U$ limit. 
Up to a mapping from the rotor operators to the spin ladder operators, $e^{\pm i \theta_i } = L^{\pm}_i $, 
this charge sector Hamiltonian is equivalent to a Kagom\'e lattice 
spin-1/2 XXZ model in the presence of an external magnetic field. We can recast the charge sector Hamiltonian as
\begin{eqnarray}
    H_{\text{c}} & = &
     \sum_{ \langle ij \rangle   } \big[ 
    {- J_{ ij}} ( L^+_i L^-_j + h.c. ) + V_{ ij} L_i^z L_j^z
    \big]
    \nonumber \\
    && + B^{\text{eff}} \sum_i L_i^z.
\end{eqnarray}
Here the effective spin-$1/2$ ladder operators $L_i^{\pm}$ satisfy
\begin{equation} L^{\pm}_i |  L_i^z = \mp \frac{1}{2} \rangle = | L_i^z
=  \pm \frac{1}{2} \rangle,
\end{equation}
and the effective ``magnetic'' field reads $B^{\text{eff}} \equiv h + 3(V_1+V_2)$. 
The $1/6$ electron filling can be regarded as the total ``magnetization'' condition
${{N_s}^{-1} \sum_i L_i^z = - {1}/{6}}$, 
where $N_s$ is the total number of Kagom\'e lattice sites. 

\begin{figure}[t]
{\includegraphics[width=8.6cm]{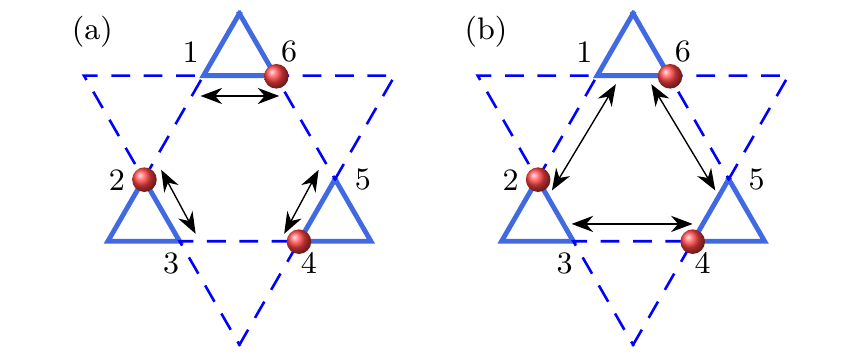}}
\caption{ 
The two collective hopping processes that contribute to the 
ring electron hopping or the ring exchange in Eq.~\eqref{Hring}. 
The red ball represents the electron or the charge rotor. 
}
\label{figS}
\end{figure}

When the inter-site repulsion $V_{1,2}$ dominate over the hoppings 
$t_{1,2}$, the system enters into the type-II CMI phase where the cluster 
localization appears on both types of triangles. In terms of the effective spin $L_i^z$, 
the electron charge localization condition in the type-II CMI is
\begin{equation}
\sum_{  i \in \text{u}} L^z_i = - \frac{1}{2},\quad
\sum_{  i \in \text{d}} L^z_i = - \frac{1}{2}. 
\label{eq08}
\end{equation}
Therefore, the allowed effective spin configuration is ``2-down 1-up'' in every triangle. 
These allowed classical spin configurations are extensively degenerate. 
But the degeneracy will be further lifted after involving the 
transverse effective spin exchanges. Physically, the effective interactions are originated 
from the collective hopping processes of electrons (shown in Fig.~\ref{figS}) and 
can be obtained from a third order degenerate perturbation theory. The resulting ring 
exchange Hamiltonian has the form of 
\begin{equation}
    H_{\text{c,ring}} = - \sum_{\hexagon} J_{\text{ring}} 
     (L^+_1 L^-_2 L^+_3 L^-_4 L^+_5 L^-_6 + h.c. ),
    \label{Hring}
\end{equation}
where ``$\hexagon$'' refers to the elementary hexagon of the 
Kagom\'e lattice,  
$J_{\text{ring}} = 
6 J_1^3 / V_2^2  + 6 J_2^3 / V_1^2$ is the ring exchange parameter and ``1, 2, 3, 4, 5, 6'' 
are the six vertices on the corner of the elementary hexagon 
on the Kagom\'e lattice (see Fig.~\ref{figS}). 

%-----------------------------
%-----------------------------

We now demonstrate that the effective Hamiltonian $H_{\text{c,ring}}$ can be mapped into
 a compact U(1) lattice gauge theory on the dual honeycomb lattice. 
As shown in Fig.~\ref{fig1}(a), this dual honeycomb lattice is 
formed by the centers of up- and down- triangles, labeled as $\bf{r}$ 
and $\bf{r}'$ respectively. We follow the previous work Ref.~\onlinecite{PhysRevB.69.064404} 
and introduce the lattice U(1) gauge fields ($E, A$) by defining
\begin{eqnarray} 
    L^z_{{\bf r},\mu} & \equiv & {  L^z_{{\bf r} + \frac{{\bf e}_{\mu}}{2}}  }= E_{ {\bf r}, {\bf r} + {\bf e}_{\mu} }, \\
    \quad L^{\pm}_{{\bf r},\mu} &\equiv& L^{\pm}_{{\bf r} + \frac{{\bf e}_{\mu}}{2}}= e^{ \pm i A_{ {\bf r},{\bf r} +{\bf e}_{\mu} }},
    \label{Gauge2}
\end{eqnarray}
where ${\bf r} \in \text{u}$, ${E_{{\bf r}{\bf r}\rq{}} = - E_{ {\bf r}\rq{} {\bf r}}}$, 
and $A_{ {\bf r}{\bf r}\rq{} } = - A_{{\bf r}\rq{}{\bf r}}$. 
The fields $E$ and $A$ are identified as the electric field and the vector gauge field 
of the compact U(1) lattice gauge theory and satisfy
$ [  E_{ {\bf r}, {\bf r} + {\bf e}_{\mu} },  A_{ {\bf r}, {\bf r} + {\bf e}_{\mu} } ] = - i$. 
With this identification, the local ``2-down 1-up'' 
charge localization condition in Eq.~\eqref{eq08} 
is interpreted as the ``Gauss' law'' 
for the emergent U(1) lattice gauge theory.
The effective ring exchange Hamiltonian $H_{\text{c,ring}}$
reduces to a gauge ``magnetic'' field term on the dual honeycomb lattice,
\begin{eqnarray}
H_{\text{c,ring}} = -2J_{\text{ring}}
\sum_{\varhexagon} \cos ( \Delta \times A ),
\label{eq12}
\end{eqnarray}
where $\Delta \times A$ is a lattice curl defined on the ``$\varhexagon$'' 
that refers to the elementary hexagon on the dual honeycomb lattice.
As this internal gauge structure emerges at the low energies
in the charge sector, we refer this gauge field as the U(1)$_{\text{c}}$ 
gauge field.  
The fate of U(1)$_{\text{c}}$ gauge field can become confining as 
it is in two spatial dimensions. However, as the gapless spinon 
matter is involved, it is likely that the instanton events can still 
get suppressed. Thus, even though the plaquette charge 
order is expected in previous works~\cite{Gang2016,Pollmann2008,Banerjee2008,Fiete2011,Ferhat2014,Pollmann2014}, 
the gauge deconfinement can coexist with the charge order. This is 
very much like the AFM$^{\ast}$ phase where
the spin quantum number fractionalization and the antiferromagnetic order
coexist~\cite{Lannert_2003,PhysRevB.64.014518}. 
The more detailed structure inside the charge sector of the type-II CMI is not the focus of this work.
We are more concerned about the localization pattern and thus assume the 
charge fractionalization in the type-II CMI. A strict analysis requires non-perturbative computations involving quantum fluctuations and is beyond the mean-field theory. In this work, we ignore the instanton effect and focus on constructing the mean-field phase diagram for the extended Hubbard model.

\section{Mean-field theory for the transition between type-I to type-II CMIs}
\label{secV}

In this section, we go beyond the strong breathing limit and build a generic framework which can support both type-I and type-II CMIs. As mentioned in the last section, the slave-rotor representation is incapable of describing the type-II CMI state where cluster localization occurs on both types of triangles. Therefore, we first introduce a new parton representation based on the emergent gauge structure in type-II CMI and then establish the phase diagram at the mean-field level. We also discuss the properties in each clusterization phase and the phase transitions between them.

\subsection{Slave-particle Construction and Mean-field Theory}

To study the transition between two distinct cluster 
localization states, we return to the charge sector Hamiltonian in Eq.~\eqref{eq3} 
by adding the $V_2$ interaction,
\begin{eqnarray}
H_{\text{c}} &=& - \sum_{\langle ij \rangle } 2{J}_{ij} \cos (\theta_i -\theta_j) 
+ \sum_{\langle ij \rangle  } V_{ij} (L_i^z +\frac{1}{2})
\nonumber \\
&& \times (L_j^z +\frac{1}{2}) + \sum_i \big[ \frac{U}{2} (L_i^z)^2  
+ h (L_i^z +\frac{1}{2})             \big]. 
\label{eq6}
\end{eqnarray}
Since the electron is not localized on a lattice site in the CMIs, 
the rotor variable $e^{i \theta_i}$ is insufficient 
to describe all the phases and phase transitions, except for
the special limits for type-I CMI that we have analyzed.  
To fix the problem, we extend the slave-rotor representation 
to a new parton construction for the electron operator~\cite{Lee2012,Savary2012,Gang2014},
\begin{equation}
c^{\dagger}_{ {\bf r}\mu\sigma} 
= f^\dagger_{ {\bf r}\mu\sigma }
{\Phi}^{\dagger}_{\bf r} {\Phi}^{\phantom\dagger}_{{\bf r} + {\bf e}_{\mu}}
l^+_{ {\bf r},{\bf r}+{\bf e}_{\mu}},
\label{SlaveP}
\end{equation}
where ${\bf e}_{\mu}$ connects the up-triangle center 
${\bf r}$ and the neighboring down-triangle centers ${{\bf r} + {\bf e}_{\mu}}$, 
${\Phi}_{\bf r}^{\dagger}$ (${\Phi}_{{\bf r}+{\bf e}_{\mu}}^{\phantom\dagger}$)
creates (annihilates) the bosonic charge excitation in the triangle 
at ${\bf r}$ (${\bf r}+{\bf e}_{\mu}$), and ${l^{\pm}_{{\bf r},{\bf r}+{\bf e}_{\mu}} 
\equiv | l^{\pm}_{{\bf r},{\bf r}+{\bf e}_{\mu}}  | e^{\pm i A_{{\bf r},{\bf r}+{\bf e}_{\mu}} }}$ 
is an open string operator of the U(1)$_{\text{c}}$ gauge field      
in the charge sector connecting the charge excitations in the 
neighboring triangles at ${\bf r}$ and ${{\bf r} + {\bf e}_{\mu}}$.  
Under the U(1)$_{\text{c}}$ gauge transformation, 
${{\Phi}^{\dagger}_{\bf r}  \rightarrow {\Phi}^{\dagger}_{\bf r}\, e^{i \chi_{\bf r}}}, 
{{\Phi}_{{\bf r} + {\bf e}_{\mu}}  \rightarrow {\Phi}_{{\bf r} + {\bf e}_{\mu}} 
 e^{- i \chi_{{\bf r} + {\bf e}_{\mu}}} }$, and $
{ A_{{\bf r},{\bf r} + {\bf e}_{\mu}}  \rightarrow 
  A_{{\bf r},{\bf r} + {\bf e}_{\mu}}   e^{-i \chi_{\bf r}
   +  i \chi_{{\bf r} + {\bf e}_{\mu}}} }$.
 To constrain the Hilbert space of the parton construction, 
 one defines the following operator~\cite{Lee2012}, 
\begin{eqnarray}
Q_{\bf r} &=&\frac{\eta_{\bf r}}{2} {+  \eta_{\bf r} } 
\sum_{\mu} L^z_{  {\bf r},{\bf r} { + \eta_{\bf r} }{\bf e}_{\mu}  } 
\nonumber  \\
&\equiv &
\frac{\eta_{\bf r}}{2} {+ \eta_{\bf r}  \sum_{\mu} }
l_{  {\bf r},{\bf r} + \eta_{\bf r} {\bf e}_{\mu}  } ,
\end{eqnarray}
that measures the local U(1)$_{\text{c}}$ (electric) gauge charge,
and for the remaining part of the paper, ${\bf r}$ refers to the centers 
of both up (denoted as `u') and down (denoted as `d') triangles.  
Here, ${\eta_{\bf r} = +1}$ ($-1$) for ${{\bf r} \in \text{u}}$ (${{\bf r} \in \text{d}}$)
and ${l_{  {\bf r},{\bf r} + \eta_{\bf r} {\bf e}_{\mu}  } {=  L^z_{  {\bf r},{\bf r} 
+ \eta_{\bf r} {\bf e}_{\mu}}}}$. 
We further supplement this definition with a Hilbert space constraint~\cite{Lee2012,Savary2012},
\begin{eqnarray}
&&[{\Phi}^{\phantom\dagger}_{\bf r}, Q^{}_{\bf r} ] = {\Phi}^{\phantom\dagger}_{\bf r},
\quad [{\Phi}^{\dagger}_{\bf r},  Q_{\bf r}^{} ] =    - {\Phi}^{\dagger}_{\bf r},
\end{eqnarray}
such that the 
physical Hilbert space is recovered. For the type-II CMI, ${Q_{\bf r} = 0}$ 
for every triangle.  

Due to the single electron occupancy on all triangles for type-II CMI, 
the electron motions are correlated in type-II CMI instead of the free electron 
motion in the inset of Fig.~\ref{fig2}(a) for a type-I CMI. This correlated
electron motion leads to the emergent U(1)$_\text{c}$ gauge structure here~\cite{Gang2016} 
and the 
plaquette charge order whose consequences on the 
spin sectors are explained in previous works~\cite{Gang2016,Pollmann2008,Banerjee2008,Fiete2011,Ferhat2014,Pollmann2014}. 
This is not the focus of this work where
we are more concerned about the distinct types of cluster localization.
Using the new parton construction, the Hubbard model becomes
\begin{eqnarray}
H &=&
 -t_1 
\sum_{{\bf r} \in \text{u}}
 \sum_{\mu\neq \nu}
l^+_{{\bf r}, {\bf r} + {\bf e}_{\mu} }  l^-_{{\bf r}, {\bf r} + {\bf e}_{\nu}} 
f^{\dagger}_{ {\bf r}\mu\sigma } f^{\phantom\dagger}_{ {\bf r}\nu\sigma} 
\Phi^{\dagger}_{ {\bf r} + {\bf e}_{\mu} } 
\Phi^{\phantom\dagger}_{ {\bf r} + {\bf e}_{\nu} }
\nonumber \\
&&
-t_2 
\sum_{{\bf r} \in \text{d}} 
\sum_{\mu\neq \nu}
l^+_{ {\bf r} - {\bf e}_{\mu},{\bf r} }  l^-_{ {\bf r} - {\bf e}_{\nu}, {\bf r}} 
f^{\dagger}_{ {\bf r}\mu\sigma } f^{\phantom\dagger}_{ {\bf r}\nu\sigma} 
\Phi^{\dagger}_{ {\bf r} - {\bf e}_{\mu} } 
\Phi^{\phantom\dagger}_{ {\bf r} - {\bf e}_{\nu} }
\nonumber \\
&& + \frac{V_1}{2} \sum_{{\bf r} \in \text{u} } Q_{\bf r}^2 
+ \frac{V_2}{2} \sum_{{\bf r} \in \text{d} } Q_{\bf r}^2,
\end{eqnarray}
which is supplemented with the Hilbert space constraint. 
In mean-field treatment, we decouple the kinetic terms and 
$H_{\text{c}}^{\text u} $ for the charge sector in the up-triangles,
$H_{\text{c}}^{\text d} $ for the charge sector in the down-triangles,
$H_{\text{s}}$ for the spinon, and $ H_{l} $ for the U(1) gauge link,
\begin{eqnarray}
&& H_{\text{c}}^{\text u} = 
\sum_{{\bf r}\in \text{d} }
\sum_{ \mu \neq \nu} {-\bar{J}_1 }
\Phi^{\dagger}_{{\bf r} - {\bf e}_{\mu}} 
\Phi^{\phantom\dagger}_{{\bf r}-{\bf e}_{\nu} } 
+ \frac{V_1}{2} \sum_{{\bf r} \in \text{u}} Q_{\bf r}^2, 
\nonumber 
\\
&& H_{\text{c}}^{\text d} = 
\sum_{{\bf r}\in \text{u} }
\sum_{ \mu \neq \nu} {-\bar{J}_2 }
\Phi^{\dagger}_{{\bf r} + {\bf e}_{\mu}} 
\Phi^{\phantom\dagger}_{{\bf r}+{\bf e}_{\nu}} 
+ \frac{V_2}{2} \sum_{{\bf r} \in \text{d}} Q_{\bf r}^2,
\nonumber
\\
&& H_{\text{s}} = 
\sum_{\mu \neq \nu} \big[ 
{ - \bar{t}_1}
  \sum_{{\bf r}\in \text{u} }
 f^{\dagger}_{{\bf r}\mu\sigma }
 f^{\phantom\dagger}_{{\bf r}\nu\sigma } 
 {-\bar{t}_2 }
 \sum_{{\bf r}\in \text{d} } 
 f^{\dagger}_{{\bf r}\mu\sigma }
 f^{\phantom\dagger}_{{\bf r}\nu\sigma } \big],
 \nonumber
 \\
&& H_{l} = \sum_{ \mu \neq \nu } \big[ {- \bar{K}_1 } 
\sum_{{\bf r} \in \text{u} } 
 l^+_{{\bf r}, {\bf r} + {\bf e}_{\mu} }  l^-_{{\bf r}, {\bf r} + {\bf e}_{\nu}} 
{- \bar{K}_2}  \sum_{{\bf r} \in \text{d} } 
 l^+_{ {\bf r} - {\bf e}_{\mu},{\bf r} }  l^-_{ {\bf r} - {\bf e}_{\nu}, {\bf r}} \big] .
 \nonumber
\end{eqnarray}
where the mean-field parameters are defined by
\begin{eqnarray}
\bar{J}_1 &=& t_2 
\langle l^+_{ {\bf r} - {\bf e}_{\mu},{\bf r} } \rangle
\langle l^-_{ {\bf r} - {\bf e}_{\nu}, {\bf r} }  \rangle
\sum_{\sigma}
\langle f^{\dagger}_{ {\bf r}\mu\sigma } f^{\phantom\dagger}_{ {\bf r}\nu\sigma } \rangle,
\quad {\bf r} \in \text{d} ,
\nonumber
\\
\bar{J}_2 &=& 
t_1
\langle l^+_{ {\bf r}, {\bf r}+{\bf e}_{\mu} } \rangle 
\langle l^-_{  {\bf r}, {\bf r}+{\bf e}_{\nu} } \rangle
\sum_{\sigma}
\langle f^{\dagger}_{ {\bf r}\mu\sigma } f^{\phantom\dagger}_{ {\bf r}\nu\sigma } \rangle,
 \quad {\bf r} \in \text{u} ,\nonumber
\\
\bar{t}_1 &=& 
t_1 \langle l^+_{ {\bf r}, {\bf r}+{\bf e}_{\mu} } \rangle  
 \langle l^-_{ {\bf r}, {\bf r}+{\bf e}_{\nu} } \rangle  
\langle  {\Phi}^{\dagger}_{{\bf r} + {\bf e}_{\mu}} 
\Phi^{\phantom\dagger}_{{\bf r}+{\bf e}_{\nu}} 
\rangle,
\quad {\bf r} \in \text{u},\nonumber
\\
\bar{t}_2 &=&
t_2
\langle l^+_{ {\bf r} - {\bf e}_{\mu},{\bf r} } \rangle
\langle l^-_{ {\bf r} - {\bf e}_{\nu}, {\bf r} }  \rangle
\langle  \Phi^{\dagger}_{{\bf r} - {\bf e}_{\mu}} 
\Phi^{\phantom\dagger}_{{\bf r} - {\bf e}_{\nu}} 
\rangle,
\quad {\bf r} \in \text{d}, \nonumber
\\
\bar{K}_1 &=& t_1
\sum_{\sigma}
\langle f^{\dagger}_{ {\bf r}\mu\sigma } 
f^{\phantom\dagger}_{ {\bf r}\nu\sigma } \rangle
\langle  \Phi^{\dagger}_{{\bf r} + {\bf e}_{\mu}} 
\Phi^{\phantom\dagger}_{{\bf r}+{\bf e}_{\nu}} 
\rangle,\quad\quad {\bf r} \in \text{u}, \nonumber
\\ 
\bar{K}_2 &=&
 t_2
\sum_{\sigma}
\langle f^{\dagger}_{ {\bf r}\mu\sigma } 
f^{\phantom\dagger}_{ {\bf r}\nu\sigma } \rangle
\langle  \Phi^{\dagger}_{{\bf r} - {\bf e}_{\mu}} 
\Phi^{\phantom\dagger}_{{\bf r}-{\bf e}_{\nu}} 
\rangle,\quad\quad {\bf r} \in \text{d}. \nonumber
\end{eqnarray}
Here we have dropped the Lagrange multipliers in the decoupled mean field Hamiltonian. 
Because they arise from the constraints of physical Hilbert space and expected to vanish for the single occupation condition within all triangles in type-II CMI. 
In the decoupling treatment, we also respect all symmetries of the original Hubbard model 
to obtain correct mean-field parameters defined above.

\subsection{Mean-field Phase Diagram}

To solve the bosonic mean-field Hamiltonians, we introduce a rotor variable $\varphi_{\bf r}$ 
that is conjugate to the U(1)$_{\text{c}}$ charge operator $Q_{\bf r}$ with
\begin{equation}
[\varphi_{\bf r}, Q_{\bf r}] = i,
\end{equation}
and hence
\begin{eqnarray}
&& \Phi_{\bf r} = e^{-i \varphi_{\bf r}},\\
&& \Phi^\dagger_{\bf r} \Phi^{\phantom\dagger}_{\bf r} =1. 
\end{eqnarray}
After carrying out the coherent state path integral for the $\Phi_{\bf r}$ fields and integrating out the $Q_{\bf r}$ field,
the resulting partition functions for the up- and down-triangles share the same form as
\begin{equation}
\mathbb{Z}_{i} = \int {\mathcal D} \Phi^\dagger {\mathcal D} 
\Phi {\mathcal D} \lambda
e^{- {\mathcal S}_{i} - \sum_{{\bf r} \in i} \int d\tau
\lambda_{\bf r} ( | \Phi_{\bf r}|^2 -1 )  }, 
\end{equation}
where $i$ can take `u' (or 1) and `d' (or 2) corresponding to up- and down-triangle subsystems respectively.
The Lagrange multiplier $\lambda_{\bf r}$ is used to implement the unimodular constraint for the 
$\Phi$ field at each ${\bf r}$ site. 
The effective action $\mathcal{S}_{i}$ for the up- and down-triangle subsystems are
\begin{equation}
\mathcal{S}_{i} = \int d\tau \sum_{{\bf r} \in i} 
\frac{1}{2V_i} | \partial_{\tau} \Phi_{\bf r}  |^2 
- \bar{J}_i \sum_{\langle {\bf r} {\bf r}' \rangle \in i} 
(
\Phi_{\bf r}^\dagger \Phi_{\bf r}^{\phantom\dagger} 
+ h.c.
),
\end{equation}
where $\langle {\bf r}{\bf r}'\rangle$ refers to the nearest-neighbor sites 
on each type of triangle subsystem.
The rest of the treatment on each subsystems is identical to what we did to the super-rotor mode in Sec.~\ref{secIII}
and then we can find the critical $V_i/\bar{J}_i$ at which the bosons 
are condensed. 
The resemblance between the above actions and the the action of 
Eq.~\eqref{eq02} indicates the close connection between this 
slave-particle approach used here and the slave-rotor formulation used in Sec.~\ref{secIII}. 
In the strong breathing limit with $V_2=0$, this two approaches should 
give qualitatively the same results. 
But quantitatively, the current approach, through the string parameters, 
takes into account of the reduction of the spinon or electron bandwidth 
due to the on-site Hubbard interaction. 
As a result, we expect that it could give a more reliable phase diagram 
especially for the FL metal phase.

The generic mean-field phase diagrams for different choices of couplings 
are depicted as Fig.~\ref{fig2}(b,c). The four phases correspond to 
different behaviors of the charge bosons (see Tab.~\ref{Tab1}). 
When the charge bosons from both up- and down-triangles are condensed,
the FL metal is realized. 
When they are both gapped and uncondensed, we have the type-II CMI. 
When the charge bosons from one triangle are condensed and the other
is uncondensed, we have the type-I CMI. 
Here the subindex `u' or `d' to the type-I CMI indicates
which triangles the electrons are localized in. 
In mean-field theory, because the charge is a higher energy degree of freedom,
the spinon sector was treated as a spectator, rather than 
the driving force.

\begin{table}[t]
\centering
\begin{tabular}{ll}
\hline\hline 
Type-II CMI \quad \quad & $\langle \Phi_{\bf r} \rangle = 0$ for ${\bf r} \in \text{u,\;d}$.
\vspace{1mm}
\\
Type-I$_{\text u}$ CMI \quad \quad   & $\langle \Phi_{\bf r} \rangle = 0$ for ${\bf r} \in \text{u}$,  $\langle \Phi_{\bf r} \rangle \neq 0$ for ${\bf r} \in \text{d}$.
\vspace{1mm}
\\
Type-I$_{\text d}$ CMI \quad \quad   & $\langle \Phi_{\bf r} \rangle \neq 0$ for ${\bf r} \in \text{u}$,  $\langle \Phi_{\bf r} \rangle = 0$ for ${\bf r} \in \text{d}$.
\vspace{1mm}
\\
FL metal &  $\langle \Phi_{\bf r} \rangle \neq 0$ for ${\bf r} \in \text{u}$,  $\langle \Phi_{\bf r} \rangle \neq 0$ for ${\bf r} \in \text{d}$.
\vspace{1mm}
\\
\hline\hline
\end{tabular}
\caption{The description of the charge sector of the four different phases 
in the slave-particle formalism.}
\label{Tab1}
\end{table}

We turn to explain the phase boundaries in Fig.~\ref{fig2}(b,c). 
As we increase $V_2/t_1$, the effective electron 
 hopping on the up-triangle bonds gets suppressed which effectively 
 enhances the kinetic energy gain through the down-triangle bonds. 
 Thus, a larger $V_1/t_2$ is required to drive a Mott transition. 
 A similar argument applies to the boundary between type-I$_{\text u}$ 
 and type-II CMIs. A larger $V_2/t_1$ is needed to compete with the kinetic energy 
 gain on the up-triangle bonds for a larger $V_1/t_2$ in type-I$_{\text u}$ 
CMI and to drive a transition to type-II CMI. For ${t_1 > t_2}$, electrons are 
more likely to be localized in the up-triangles to gain the intra-cluster kinetic energy. 
Thus, a smaller $V_1/t_2$ is needed to drive a Mott transition and 
a larger $V_2$ is needed to drive the system 
from type-I$_{\text u}$ to type-II CMIs. 

The phase transition between the tpye-I and type-II CMIs can also be understood in the charge boson picture. To be concrete, we focus on the transition from the type-II CMI to 
the tpye-I$_{\text u}$ CMI and the extension to the tpye-I$_{\text d}$ CMI is direct. 
In the type-II CMI, charge bosons from both up- and down-triangles 
are gapped and uncondensed. 
With the decreasing of the inter-site repulsion $V_2/t_1$ on the down-triangle bonds,
the charge bosons become condensed on the down-triangle subsystem 
and the U(1)$_{\text{s}}$ gauge field also acquires a mass concurrently. 
The two fractionally-charged charge bosons $\Phi$ from two types of triangles
then are combined back into the original unit-charged charge rotor $e^{i \theta}$. 
The large inter-site repulsion $V_1/t_t$  still preserves 
the single electron occupancy on the up-triangle subsystem. 
Thus, the charge rotor $e^{i \theta}$ is well-defined on the 
center of an up-triangle and within the up-triangle, the localized electron can 
move more or less freely. 
In this sense, the condensation of the charge bosons from the down-triangles 
leads to the local ``metallic'' clusters in the up-triangles. 
After the charge boson condensation, there is no charge fractionalization 
in the type-I CMI, but the spin-charge separation still survives. 
Because of the local ``metallic'' clusters, only the U(1)$_{\text{s}}$ gauge field 
living on the down-triangle bonds that connect the up-triangles remains active 
and continues to fluctuate at the low energies. 
The low-energy physics is described by the spinon Fermi surface coupled 
with a fluctuating U(1)$_{\text{s}}$ gauge field, leading to a U(1) QSL 
in the triangular lattice formed by the up-triangles.

All transitions discussed above are continuous at 
mean-field level, except the transition between type-I$_{\text u}$ and  
type-I$_{\text{d}}$ CMIs that is strongly first order. Beyond mean-field theory, 
the transition between FL metal and type-I CMIs will remain
continuous and quantum XY type~\cite{Senthil2008A,Senthil2008B} 
while the transition into type-II CMI 
may depend on the detailed charge structure inside type-II CMI. Moreover, our mean field theory does not capture the charge 
quantum fluctuation inside the type-II CMI as described by the compact 
U(1)$_{\text{c}}$ gauge theory in Sec.~\ref{secIV}, but does  
obtain qualitatively correct phase boundaries.

\section{Discussion}
\label{secVI}

We discuss the experimental relevance and consequences about the
Mo-based cluster magnets. These compounds, \ce{M2Mo3O8} 
(M = Mg, Mn, Fe, Co, Ni, Zn, Cd), \ce{LiRMo3O8} (R = rare earth) 
and other related variants~\cite{Gall2013,McCarroll1957,Torardi1985,McCarroll1977}, 
incorporate the \ce{Mo3O13} cluster unit, 
and the physical properties of most materials have not been carefully 
studied so far. According to our theory, more anisotropic systems with 
a stronger breathing tend to favor the type-I CMI. \ce{Li2InMo3O8} 
is more anisotropic than \ce{LiZn2Mo3O8} from the lattice parameters. 
For \ce{LiZn2Mo3O8}, the spin susceptibility shows a ``1/3 anomaly'' 
and double Curie regimes~\cite{Sheckelton2012,McQueen2014}. 
This was attributed to the plaquette charge order
in the type-II CMI that reconstructs the spin sector. In contrast, \ce{Li2InMo3O8}
is characterized by one Curie regime with the Curie
temperature ${\Theta_{\text{CW}} = -207}$K down to 25K~\cite{Gall2013}. 
The Curie constant is consistent with one unpaired 
spin-1/2 moment per \ce{Mo3O13} cluster in the type-I CMI. 
Below 25K, the spin susceptibility of \ce{Li2InMo3O8} 
saturates to a constant, which is consistent with the expectation from
a spinon Fermi surface U(1) QSL. Besides the structural and spin 
susceptibility data, however, very little is known about \ce{Li2InMo3O8}. 
It is also likely that this system is located in the 120-degree order state of the model. 
Thus, more experiments are needed to confirm the absence of magnetic ordering 
in \ce{Li2InMo3O8} and also to explore the magnetic properties of
\ce{ScZnMo3O8} and other cluster magnets~\cite{Haraguchi2015}.
From the numerical aspect, first principal calculation is carried out recently and 
supports the experimental findings and theoretical understandings in \ce{Mo3O8} magnets.
Namely, \ce{LiZn_2Mo3O8} is shown to exhibit the plaquette order with one dangling spin,
\ce{Li2InMo3O8} is a CMI with 120-degree order and \ce{Li2ScMo3O8} displays 
a spin liquid behavior\cite{Streltsov2020}.

\begin{acknowledgments}
We thank the conversation about twisted bilayer graphene with 
Jianpeng Liu. XQW is supported by MOST 2016YFA0300501, NSFC 11974244,
 and from a Shanghai talent program. YBK is supported by NSERC of Canada 
 and the Killam Research Fellowship from the Canada Council of the Arts.
The remaining authors are supported by the Ministry of Science and Technology of China with 
Grant No.~2018YFGH000095, 2016YFA0301001, 2016YFA0300500, by Shanghai Municipal 
Science and Technology Major Project with Grant No.2019SHZDZX04, and by the Research 
Grants Council of Hong Kong with General Research Fund Grant No.17303819 and
No. 17306520. 
\end{acknowledgments}

\bibliography{ref}

\end{document}